\documentclass[pra,aps,twocolumn,superscriptaddress,nofootinbib,a4paper]{revtex4}

\usepackage{latexsym}
\usepackage{hyperref}
\usepackage{epsfig}
\usepackage{psfrag}
\usepackage{graphics}
\usepackage{amsmath}
\usepackage{xspace}
\usepackage{amssymb}
\usepackage{amsthm}

\def\uin{|\!\hspace{0.45mm}|\!\hspace{0.45mm}|}
\theoremstyle{plain}
\newtheorem{theorem}{Theorem}
\newtheorem{proposition}{Proposition}

\theoremstyle{definition}

\theoremstyle{remark}

\begin{document}

\title{Frustration, interaction strength, and ground-state
  entanglement in complex quantum systems}

\author{Christopher~M.~Dawson}
\email{dawson@physics.uq.edu.au}
\affiliation{School of Physical Sciences, The University of Queensland, Brisbane Queensland 4072, Australia}
\affiliation{Institute for Quantum Information, California Institute of Technology, Pasadena CA 91125, USA}
\author{Michael~A.~Nielsen} 
\email{nielsen@physics.uq.edu.au}
\homepage{www.qinfo.org/people/nielsen} 
\affiliation{School of Physical Sciences, The University of Queensland, Brisbane Queensland 4072, Australia}
\affiliation{Institute for Quantum Information, California Institute of Technology, Pasadena CA 91125, USA}
\affiliation{School of Information Technology and Electrical Engineering, The University of Queensland, Brisbane Queensland 4072, Australia}


\date{\today}

\begin{abstract}
  Entanglement in the ground state of a many-body quantum system may
  arise when the local terms in the system Hamiltonian fail to commute
  with the interaction terms in the Hamiltonian.  We quantify this
  phenomenon, demonstrating an analogy between ground-state
  entanglement and the phenomenon of frustration in spin systems.  In
  particular, we prove that the amount of ground-state entanglement is
  bounded above by a measure of the extent to which interactions
  \emph{frustrate} the local terms in the Hamiltonian.  As a
  corollary, we show that the amount of ground-state entanglement is
  bounded above by a ratio between parameters characterizing the
  strength of interactions in the system, and the local energy scale.
  Finally, we prove a qualitatively similar result for other energy
  eigenstates of the system.
\end{abstract}
\maketitle

\section{Introduction}

%
%
A central problem in physics is understanding the ground-state
properties of a complex many-body Hamiltonian, especially the
ground-state correlations.  As an outgrowth of that interest, there
has recently been considerable work on understanding the
\emph{non-classical} correlations in the ground state, that is, the
\emph{ground-state entanglement}.  Some recent work on this problem,
with further references,
includes~\cite{Haselgrove03c,Haselgrove03b,Vidal03a,Latorre03a,Tessier03a,Costi03a,Hines03a,Osterloh02a,Osborne02a,Scheel02a,Wang02b,Oconnor01a,Gunlycke01a,Nielsen98d,Jordan03a}.
This work has been motivated by the remarkable recent progress in
using entanglement as a physical resource to accomplish feats such as
quantum computation and quantum
teleportation\footnote{See~\cite{Nielsen00a,Preskill98c} for reviews
  and further references.}.

%
%
\begin{figure}[t]
\epsfig{file=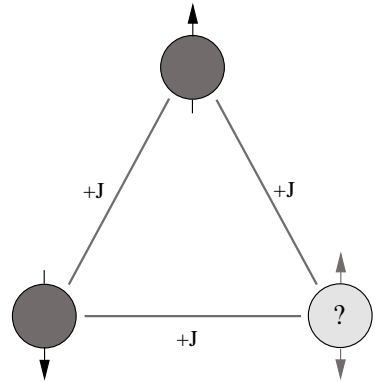,height=2in}
\caption{A system containing three spin-$\frac 12$ particles,
  coupled by a classical anti-ferromagnetic coupling ($+J\sigma_z
  \sigma_z$, with positive coupling strength $J$) favouring
  anti-alignment.  There is no way all the competing couplings can be
  simultaneously satisfied; for this reason we say the system is
  \emph{frustrated}.
\label{fig:frustration}}
\end{figure}

%
%
In this paper we connect the phenomenon of ground-state entanglement
to a well-known idea in condensed matter physics, that of
\emph{frustration}, which we now briefly review.  More detailed
introductions may be found in~\cite{Aeppli97a}.  A typical example of
a frustrated spin system is shown in Fig.~\ref{fig:frustration}.  It
consists of a triangular arrangement of three spin-$\frac 12$
particles, each pair being coupled by a classical anti-ferromagnetic
coupling ($+J\sigma_z \sigma_z$, with positive coupling strength $J$).
The anti-ferromagnetic coupling means that neighbours prefer to be
anti-aligned in order to minimize their interaction energies.  However,
a little thought shows that it is impossible for all three spins to
simultaneously be anti-aligned with each of their neighbours. It is
therefore not possible to simultaneously minimize all three
interaction energies, and the system is said to be \emph{frustrated}
for this reason.  The ground state of the Hamiltonian is a compromise
between the minimum energy states of the interaction terms.


%
%
Let us consider an analogous example in which frustration arises not
from the difficulty of choosing simultaneously compatible spin
configurations, but rather from choosing simultaneously compatible
\emph{bases} for Hilbert space.  For example, consider a system of two
spin-$\frac 12$ particles with Hamiltonian $H = -g \left( \sigma^1_x +
  \sigma^2_x \right) - \sigma^1_z \sigma^2_z$, where the superscripts
indicate which spin the operators act on, and $\sigma_x, \sigma_y$,
and $\sigma_z$ are the usual Pauli spin operators.  The ground state
of this system arises as the result of a competitive process between
minimizing the contribution to the energy from the local Hamiltonian,
$-g \left(\sigma^1_x+\sigma^2_x\right)$, and from the interaction
Hamiltonian, $-\sigma^1_z \sigma^2_z$.  Of course, because these two
Hamiltonians do not have common eigenvectors, the actual ground state
cannot possibly minimize both simultaneously, and must be a compromise
between the respective ground states of the local and interaction
Hamiltonians.

%
%
This example suggests a connection between the ground-state
entanglement and a generalized concept of frustration.  If the
interaction term in the Hamiltonian were turned off, the system would
sit in an unentangled state --- the ground state of the local
Hamiltonian.  As the interaction term is turned on, it causes the
local Hamiltonian to become frustrated.  As a result, the ground state
sits in a basis which is a compromise between the unentangled basis of
the local Hamiltonian, and the basis for the interaction Hamiltonian.
Provided the interaction was chosen appropriately, the result will be
an entangled ground state.  Furthermore, it is clear that the more
frustrated the local Hamiltonian is by the interaction, the greater
the potential entanglement in the ground state.

%
%
The main result of this paper is a bound that makes these intuitive
ideas quantitatively precise.  Our paper thus illustrates a general
idea discussed
in~\cite{Nielsen02e,Haselgrove03c,Vidal03b,Osborne02b,Preskill00a,Nielsen98d},
namely, that quantum information science provides tools and
perspectives for understanding the properties of complex quantum
systems, complementary to the existing tools of quantum many-body
physics.

%
%
We begin in Sec.~\ref{sec:background} by reviewing some basic material
on quantitative measures of entanglement.  In Sec.~\ref{sec:main} we
prove a general, non-perturbative bound on the ground-state
entanglement, relating it to the extent to which the interaction
Hamiltonian frustrates the local Hamiltonian.  We call this the
``entanglement-frustration'' bound.  The proof of the bound is
conceptually and mathematically extremely simple.  Its interest
lies in illustrating quantitatively a connection between two
apparently disparate physical phenomena, and in the
consequences which follow from this connection, to be discussed
in later sections.

%
%
In Sec.~\ref{sec:examples} we apply the entanglement-frustration bound
to an illustrative example. Using this example, we determine necessary
conditions for the bound to saturate the ground-state entanglement. It
is then shown by construction that it is possible to come arbitrarily
close to saturation for all possible values of the ground-state
entanglement, and we conclude that the entanglement-frustration bound
is thus the strongest possible bound of its type.

%
%
Aside from its intuitive appeal and immediate relevance, the
entanglement-frustration bound has an elegant corollary described in
Sec.~\ref{sec:ent-ratio}.  Intuitively, it is clear that the
ground-state entanglement of a Hamiltonian $H=H_L+H_I$ is small if the
size of the interaction $H_I$ is small compared with some appropriate
local energy scale associated with $H_L$.  Indeed, it is
straightforward to use perturbation theory to demonstrate a bound
along these lines, valid in the limit when $H_I$ is a small
perturbation.  The entanglement-frustration bound allows us to prove a
general non-perturbative bound quantifying this intuition. This
corollary is proved in Sec.~\ref{sec:ent-ratio}.
Sec.~\ref{sec:higher-energies} generalizes these results so that they
apply to \emph{arbitrary} eigenstates of the Hamiltonian, not just the
ground state.  This is done using methods quite different from those
used in Sec.~\ref{sec:ent-ratio}, using a variant on a powerful
theorem from linear algebra known as the Davis-Kahan theorem.

%
%
The results in Secs.~\ref{sec:main}-\ref{sec:ent-ratio} provide a
compelling picture of how ground-state entanglement arises as the
result of frustration between competing local and interaction terms in
the system Hamiltonian.  Sec.~\ref{sec:higher-energies} generalizes
some of these results to apply to other energy eigenstates as well.
The paper concludes in Sec.~\ref{sec:conclusion} with a discussion of
some possible extensions to this work.

\section{Background on entanglement measures}
\label{sec:background}

%
%
To make our ideas precise we must introduce a quantitative measure of
the amount of entanglement in the ground state of a quantum system.  A
major focus of research in quantum information science over the past
few years has been developing such a theory of
entanglement\footnote{See, e.g.,~\cite{Horodecki01a,Terhal02a} for an
  introduction and further references on the theory of entanglement.},
and several good candidate measures exist.  We shall use a measure of
entanglement introduced in~\cite{Vedral97a,Vedral98a}.  For an
$n$-body quantum system in a state $\psi$ this entanglement measure is
defined by\footnote{Note that this measure is a slightly rescaled
  version of that in~\cite{Vedral97a,Vedral98a}, but has essentially
  the same properties.  In the present context the rescaled definition
  turns out to be easier to work with.}:
\begin{eqnarray} \label{eq:entanglement-measure}
  E(\psi) \equiv 1-\max_{\psi_1,\ldots,\psi_n} |\langle \psi| \psi_1 \otimes
  \ldots \otimes \psi_n\rangle|^2.
\end{eqnarray}
That is, $E(\psi)$ measures the maximal overlap $\psi$ has with a
product state $\psi_1 \otimes \ldots \otimes \psi_n$ of the $n$ bodies
making up the system.

%
%
What makes $E(\psi)$ a good entanglement measure?
\cite{Vedral97a,Vedral98a} investigated the properties of $E(\psi)$
and found that it has many properties that make it a good measure of
entanglement.  These properties include the fact that: (a) $E(\psi)$
can only decrease, never increase, under local operations and
classical communication, i.e., it is an entanglement monotone; and (b)
$E(\psi)$ is zero if and only if $\psi$ is unentangled, and otherwise
is positive.  In addition, an interesting connection has been
found~\cite{Biham02a} between $E(\psi)$ and the theory of quantum
algorithms, with $E(\psi)$ being related to the probability of success
of an algorithm whose initial state is equivalent to $\psi$, up to a
local unitary transformation.

\section{The entanglement-frustration bound}
\label{sec:main}

%
%
The general scenario we consider is an $n$-body quantum system with
Hamiltonian $H = H_L + H_I$. $H_L$ is a \emph{local Hamiltonian}
consisting of single-body or \emph{local} terms, and therefore has an
eigenbasis of unentangled states. $H_I$ contains all the remaining
terms in the Hamiltonian, and is called the \emph{interaction
  Hamiltonian}.

%
%
We let $E_0$ be the global ground-state energy, i.e., the ground state
energy of $H$, with $|E_0\rangle$ any corresponding ground state.
Similarly $E_0^L$ and $E_0^I$ are defined to be the local and
interaction ground-state energies, respectively, for $H_L$ and $H_I$.
We define the \emph{frustration energy} of the system as $E_f \equiv
E_0-E_0^L-E_0^I$.  The frustration energy thus measures the extent to
which the global ground state fails to simultaneously minimize the
local and interaction energies. It is easily shown from matrix
eigenvalue inequalities that $E_0 \geq E^L_0 + E^I_0$, so $E_f$ is
always a non-negative quantity, and is equal to zero if and only if
$H_L$ and $H_I$ have a common ground state.

%
%
Our aim is to relate the amount of entanglement in the ground state,
$E(|E_0\rangle)$, to the frustration energy, $E_f$.  Of course, to
relate the dimensionless quantity $E(|E_0\rangle)$ to $E_f$, which has
units of energy, we require another energy scale in the system.  The
relevant energy scale turns out to be associated with local
excitations of the system.  Suppose we decompose $H_L$ as
$H_1+H_2+\ldots + H_n$, where $H_j$ is the contribution to the local
Hamiltonian from the $j$th system.  Let $\Delta E_j$ be the gap
between the ground and first excited energies\footnote{\emph{Note for
    clarity:} The term ``gap'' may be used in two different senses.
  Sometimes it means the energy difference between the ground state
  and the first excited state with a strictly higher energy.  We use
  the term in the other sense, where the gap $\Delta E_j$ is zero if
  $H_j$ has a degenerate ground state.}  of $H_j$.  Let $\Delta E_{\rm
  ent}$ be the \emph{second smallest} of these energies.  That is,
suppose we choose $j_0,j_1,\ldots$ such that $\Delta E_{j_0} \leq
\Delta E_{j_1} \leq \ldots$.  Then $\Delta E_{\rm ent} = \Delta
E_{j_1}$.

%
%
Physically, $\Delta E_{\rm ent}$ is the energy we need to put into a
system with Hamiltonian $H_L$ in order to cause an excitation from the
ground state into an excited state of either system $j_0$ \emph{or}
system $j_1$.  It is thus the minimal amount of energy that we would
need to put into the system in order to cause entanglement in the
ground state, since merely exciting one system, while leaving the
others alone, leaves the system still in a product state.

%
%
Our result relating the ground-state entanglement to the frustration
energy and $\Delta E_{\rm ent}$ is the inequality:
\begin{eqnarray} \label{eq:basic-inequality}
  E(|E_0\rangle) \leq \frac{E_f}{\Delta E_{\rm ent}}.
\end{eqnarray}
We call this the \emph{entanglement-frustration} bound.  This bound
tells us that when the frustration energy is small compared with
$\Delta E_{\rm ent}$, there can't possibly be much entanglement in the
ground state of the system.  Thus, it is only systems in which the
interaction and local terms substantially frustrate one another that
it is possible to have a highly entangled ground state.

%
%
The first step in the proof of the entanglement-frustration bound,
Eq.~(\ref{eq:basic-inequality}), is to prove that
\begin{eqnarray} \label{eq:step-1}
  \langle E_0|H_L|E_0\rangle -E_0^L \leq E_f.
\end{eqnarray}
Physically, this is just the obvious statement that the extent to
which the local Hamiltonian is frustrated is no larger than the total
frustration in the system.  The proof is simply to split the
frustration energy into a sum of contributions from the local and
interaction frustration energies:
\begin{eqnarray}
  E_f & = & \langle E_0|H|E_0\rangle-E_0^L-E_0^I \\
  & = & \left( \langle E_0|H_L|E_0\rangle - E_0^L \right) + \left(
    \langle E_0 |H_I|E_0\rangle -E_0^I\right). \nonumber \\
 & & \label{eq:frustration-split}
\end{eqnarray}
The inequality of Eq.~(\ref{eq:step-1}) now follows from the
observation that $\langle E_0|H_I|E_0\rangle \geq E_0^I$.

%
%
The second step in the proof of the entanglement-frustration bound is
to expand $|E_0\rangle$ in terms of the eigenstates $|E_j^L\rangle$ of
$H_L$, $|E_0\rangle = \sum_j \alpha_j |E_j^L\rangle$.  We assume that
the local energies are ordered so that $E_0^L \leq E_1^L \leq \ldots$.
We now split the expansion of $|E_0\rangle$ into terms with energies
below $E_0^L+\Delta E_{\rm ent}$, and into terms with energies at
least $E_0^L+\Delta E_{\rm ent}$, that is,
\begin{eqnarray} \label{eq:step-2-inter}
  |E_0\rangle = \sum_{j=0}^k \alpha_j |E_j^L\rangle + \gamma |E_\perp\rangle,
\end{eqnarray}
where (a) $k$ is the largest integer such that $E_k^L < E_0^L+\Delta
E_{\rm ent}$, and thus $E_{k+1}^L = E_0^L+\Delta E_{\rm ent}$; (b)
$|E_\perp\rangle$ is a normalized state containing all the terms of
energy at least $E_0^L+\Delta E_{\rm ent}$, and thus is orthogonal to
the lower energy terms; and (c) $\gamma$ is the amplitude for
$|E_\perp\rangle$, and thus satisfies $|\gamma|^2 = 1-\sum_{j=0}^k
|\alpha_j|^2$.

For later use it is important to note that $\sum_{j=0}^{k} \alpha_j
|E^L_j\rangle$ is a product state, as all the terms $|E^L_j\rangle$
involve excitations of the \emph{same subsystem}\footnote{System
  $j_0$, to return to the notation used earlier in defining $\Delta
  E_{\rm ent}$.}. Furthermore its overlap squared with $|E_0\rangle$
is given by $\sum_{j=0}^k |\alpha_j|^2$.

Returning to the main line of the proof, from
Eq.~(\ref{eq:step-2-inter}) we have
\begin{eqnarray}
  \langle E_0|H_L|E_0\rangle = \sum_{j=0}^k |\alpha_j|^2 E_j^L +
  |\gamma|^2 \langle E_\perp|H_L|E_\perp\rangle.
\end{eqnarray}
But $E_j^L \geq E_0^L$, $\langle E_\perp|H_L|E_\perp\rangle \geq
E_0^L+\Delta E_{\rm ent}$, and $|\gamma|^2 = 1-\sum_{j=0}^k
|\alpha_j|^2$, so
\begin{eqnarray}
  \langle E_0|H_L|E_0\rangle & \geq & \sum_{j=0}^k |\alpha_j|^2 E_0^L \nonumber
\\
 & & +
  \left(1-\sum_{j=0}^k |\alpha_j|^2\right) \left(E_0^L+\Delta E_{\rm ent}
    \right). \nonumber \\
 & & 
\end{eqnarray}
Rearrangement of this inequality gives
\begin{eqnarray} \label{eq:step-2}
  \langle E_0|H_L|E_0\rangle-E_0^L \geq \left(1-
    \sum_{j=0}^k |\alpha_j|^2\right) \Delta E_{\rm ent}.
\end{eqnarray}

%
%
Combining Eqs.~(\ref{eq:step-1}) and~(\ref{eq:step-2}) we have
\begin{eqnarray}
  \left(1-\sum_{j=0}^k |\alpha_j|^2\right) \leq \frac{E_f}{\Delta E_{\rm ent}}.
\end{eqnarray}
Our desired result, Eq.~(\ref{eq:basic-inequality}), will follow if we
can establish that $E(|E_0\rangle) \leq \left(1-\sum_{j=0}^k
  |\alpha_j|^2\right)$.  This follows immediately from the definition
of the entanglement measure, Eq.~(\ref{eq:entanglement-measure}), and
the observation we made earlier in the proof, that $|E_0\rangle$ and
the product state $\sum_{j=0}^k \alpha_j |E_j^L\rangle$ have overlap
squared $\sum_{j=0}^k |\alpha_j|^2$.

\section{Application and saturation of the entanglement-frustration bound}
\label{sec:examples}

%
%
In this section we consider two separate but related issues.  First,
in Sec.~\ref{subsec:example} we apply the entanglement-frustration
bound to an illustrative and physically relevant Hamiltonian, the
two-spin transverse Ising model.  This example is used to develop
insight into the question of when the entanglement-frustration bound
is saturated.  Building on these insights, we analyse this question in
more generality in Sec.~\ref{subsec:saturation}, showing that the
entanglement-frustration bound can be saturated for all possible
values of the ground-state entanglement.  Thus, there is a sense in
which the entanglement-frustration bound is the best possible bound of
its type.


\subsection{The two-spin transverse Ising model}
\label{subsec:example}

%
%
As an illustrative example, consider a system of two
spin-$\frac{1}{2}$ particles evolving under a transverse Ising
Hamiltonian
\begin{equation}
    \label{eq:ising-ham}
    H = -g(\sigma_x^1 + \sigma_x^2) - \sigma^1_z\sigma^2_z
\end{equation}
In this model, the two particles are coupled magnetically along their
$z$ axes, and interact with an external magnetic field of strength $g$
directed along the $x$ axis. For the purposes of this example we take
$g \geq 0$.  The $g < 0$ analysis is similar, but it simplifies the
discussion to pick a definite value for the sign of $g$.

%
%
Note that while the two-spin transverse Ising model is mathematically
rather trivial, it has genuine physical interest.  Furthermore, we
will find that it is surprisingly informative as a way of
understanding the conditions under which the entanglement-frustration
bound is saturated. For these reasons we describe the results in some
detail.

%
%
Physically, $g \rightarrow 0$ is the strong coupling limit, where we
expect the ground state to become quite entangled.  We will see in
detail below that it becomes maximally entangled in this limit, i.e.,
$E(|E_0\rangle) \rightarrow \frac 12$, for our entanglement measure.
In contrast, $g \rightarrow \infty$ is the weak coupling limit, and we
expect the ground state should be a product state in that limit,
$E(|E_0\rangle) \rightarrow 0$.

%
%
The ground-state energy of~(\ref{eq:ising-ham}) is easily found to be
$E_0 = -\sqrt{1+4g^2}$, and the ground-state is
\begin{equation} \label{eq:ising-gs}
    |E_0\rangle = \frac{1}{\sqrt{N}} \left( \left(2g + \sqrt{1+4g^2}\right) 
    |++\rangle + |--\rangle \right)
\end{equation}
where $N = 1 + (2g + \sqrt{1+4g})^2$ is a normalization constant, and
$|\pm\rangle \equiv (|0\rangle\pm|1\rangle)/\sqrt 2$. Note that
$|E_0\rangle$ is in its Schmidt form, with largest Schmidt
coefficient\footnote{By contrast, if $g < 0$ the largest Schmidt
  coefficient is $\lambda_0 = 1/\sqrt{N}$.  This is the main
  difference between the $g<0$ and $g \geq 0$ cases.} $\lambda_0 =
\left(2g + \sqrt{1+4g^2}\right) / \sqrt{N}$.  The ground-state
entanglement is given by $1 - \lambda_0^2$, which simplifies to
\begin{equation}
    \label{eq:ising-gse}
    E(|E_0\rangle) = \frac{1}{2}-\frac{g}{\sqrt{1+4g^2}}.
\end{equation}

%
%
To calculate the entanglement-frustration bound we must first split
the Hamiltonian into a local and interaction part, $H_L =
-g(\sigma_x^1 + \sigma_x^2)$, and $H_I = - \sigma^1_z\sigma^2_z$.
With these choices we find that $E^L_0 = -2g$ and $E_0^I = -1$. The
two spin systems each have the same local energy spectrum with the gap
between the ground and excited states being $2g$, so we have $\Delta
E_{\mathrm{ent}} = 2g$. This gives the entanglement-frustration bound
\begin{equation}
    \label{eq:ising-fb}
    \frac{E_f}{\Delta E_\mathrm{ent}} = \frac{1+2g - \sqrt{1+4g^2}}{2g}.
\end{equation}

%
%

\begin{figure}[t]
\epsfig{file=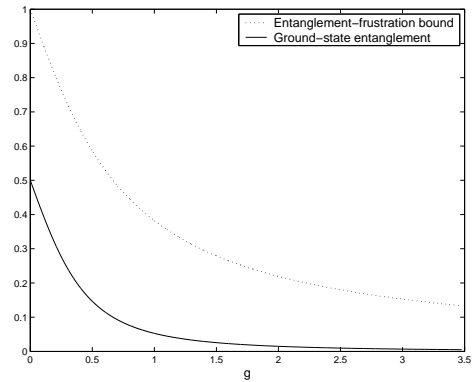,height=2in}
\caption{The ground-state entanglement and entanglement-frustration bound for the transverse Ising Hamiltonian (\ref{eq:ising-ham}) plotted against the parameter $g$. }
\label{fig:ising}
\end{figure}

A comparison of the quantities appearing in Eqs.~(\ref{eq:ising-gse})
and~(\ref{eq:ising-fb}) is shown in Fig.~\ref{fig:ising}. Both the
ground-state entanglement and the entanglement-frustration bound
decrease sharply as $g$ increases from $0$. For these small values of
$g$ the bound is approximately double the entanglement. As $g$
increases further the ground-state entanglement decreases rapidly to
$0$, while the bound decreases to $0$ more slowly. The
entanglement-frustration bound is clearly not very tight in this case,
although the qualitative behaviour of the bound and the actual
ground-state entanglement is similar.

%

We can identify two reasons for the failure to saturate the
entanglement-frustration bound in this example. First, in the language
of Sec.~\ref{sec:main}, the quantity $\langle E_\perp | H_L |E_\perp
\rangle$ is strictly larger than $E^L_0 + \Delta E_\mathrm{ent}$.  We
see from Eq.~(\ref{eq:ising-gs}) that $|E_\perp\rangle = |--\rangle$,
and thus $\langle E_\perp | H_L |E_\perp \rangle = E^L_0 + 2\Delta
E_\mathrm{ent}$.  It follows that
\begin{equation}
    \langle E_0 | H_L | E_0 \rangle = 
   \lambda_0^2 E^L_0 + (1-\lambda_0^2) (E^L_0 + 2 \Delta E_\mathrm{ent})
\end{equation}
and upon substitution into Eq.~(\ref{eq:frustration-split}) this gives
\begin{equation}
    E_f = 2 \Delta E_\mathrm{ent} (1-\lambda_0^2) + 
    \langle E_0 | H_I | E_0 \rangle - E^I_0
\end{equation}
or
\begin{equation}
    \label{eq:bound-excess}
    \frac{E_f}{\Delta E_\mathrm{ent}} =  2 E(|E_0\rangle) 
        + \frac{\langle E_0 | H_I | E_0 \rangle - 
          E^I_0}{\Delta E_\mathrm{ent}}.
\end{equation}
The entanglement-frustration bound is therefore at least twice the
ground-state entanglement with this choice of $H_L$, for all values of
$g$.


The second contribution to the excess is the term $(\langle E_0 | H_I
| E_0 \rangle - E^I_0)/ \Delta E_\mathrm{ent}$.  Physically, this is
the ratio of the frustration of the interaction energy to the local
energy scale.
The excess sharply increases from $0$ for small $g$, and decreases
slowly as $g \rightarrow \infty$. For $g$ greater than about $2$ the
ground-state entanglement is close to $0$ and the
entanglement-frustration bound is composed almost entirely of this
excess term.

%

\subsection{Saturation of the entanglement-frustration bound}
\label{subsec:saturation}

When, if ever, is the entanglement-frustration bound saturated?  We
will show in this section that for all possible values of
$E(|E_0\rangle)$ we can find a Hamiltonian $H$ whose ground state has
that amount of entanglement, and saturates the
entanglement-frustration bound as closely as desired.

%
%
Interestingly, it turns out that it is not possible to \emph{exactly}
saturate the entanglement-frustration bound except in the extreme
cases $E(|E_0\rangle) = 0$ and $E(|E_0\rangle) = 1$.  However, as we
show in this section, it is always possible to saturate the bound to
as good an approximation as desired.

%
%
To see that exact saturation is not possible, consider the necessary
condition for saturation $\langle E_0 | H_I | E_0 \rangle = E^I_0$
identified in the previous section. This condition implies that
$|E_0\rangle$ is a ground state of $H_I$, and therefore also an
eigenstate of $H_L = H - H_I$.  Entanglement in an eigenstate of a
local Hamiltonian is only possible if there is an associated
degeneracy. If $|E_0\rangle$ is a ground state of $H_L$ then we
conclude that $\Delta E_\mathrm{ent} = 0$, the
entanglement-frustration bound is undefined, and so saturation
certainly does not occur. On the other hand, if $|E_0\rangle$ is an
excited state of $H_L$ corresponding to some eigenvalue $E^L_j$ then
\begin{equation}
    \langle E_0 | H_L | E_0 \rangle - E^L_0 = E^L_j - E^L_0.
\end{equation}
But since $|E_0\rangle$ is entangled, by assumption, we must have
$E^L_j - E^L_0 \geq \Delta E_\mathrm{ent}$.  Combining this with the
result $E_f \geq \langle E_0 | H_L | E_0 \rangle - E^L_0$ gives $E_f /
\Delta E_\mathrm{ent} \geq 1$. In contrast the maximum values of
$E(|E_0\rangle)$ for qubits is $\frac{1}{2}$, and more generally for
pairs of $d$-dimensional systems it is $1 - \frac{1}{d}$. We conclude
that it is not possible for the entanglement-frustration bound to
exactly saturate, except when $E(|E_0\rangle) = 0$ or $1$.

The above analysis, however, says nothing for $|E_0\rangle$
arbitrarily close to a ground-state of $H_I$, and in these cases it is
possible that the bound approaches saturation. 

%
%
Before dealing directly with the issue of saturation, it is helpful to
address another issue, the question of a how a given many-body
Hamiltonian $H$ is to be split into local and interaction parts.
Consider, for example, the transverse Ising Hamiltonian $H =
-g(\sigma_x^1+\sigma_x^2) -\sigma_z^1 \sigma_z^2$.  In our earlier
analysis we set $H_L = -g(\sigma_x^1+\sigma_x^2)$ and $H_I =
-\sigma_z^1 \sigma_z^2$.

%
%
However, there is a certain arbitrariness in the splitting into local
and interaction Hamiltonians.  From a mathematical point of view,
there is nothing to stop us from splitting $H$ up as $H = H_L' +
H_I'$, where $H_L'$ is \emph{any} desired local Hamiltonian, and we
simply choose $H_I' \equiv H-H_L'$.  So, for example, we could choose
$H_L' = -g \sigma_x^1$ and $H_I' = -g \sigma^2_x - \sigma_z^1
\sigma_z^2$.  The reason for this ambiguity is that while the class of
local Hamiltonians is perfectly well-defined, there is no similar
definition of what it means for a Hamiltonian to be an interaction
Hamiltonian.  Failing to have such a definition, we are free to
choose $H_L$ however we like, compensating by choosing an appropriate
interaction Hamiltonian.

%
%
This freedom to choose a splitting into local and interaction parts is
reflected in the fact that the entanglement-frustration bound holds
for any choice of splitting $H = H_L + H_I$.  Of course, while
$E(|E_0\rangle)$ is not affected by the splitting chosen, the
quantities $\Delta E_{\rm ent}$ and $E_f$ are.  As a result the exact
value of the entanglement-frustration bound depends on the particular
splitting chosen.  We will use this freedom in choosing a splitting to
engineer saturation in the entanglement-frustration bound.

%
%
Physically, of course, there is often a reason to favour one splitting
into local and interaction parts over another.  For example, if we
regard the transverse Ising Hamiltonian as a model of two magnetically
coupled spins placed in an external magnetic field, then there is a
clearly-defined physical sense in which $-g(\sigma_x^1+\sigma_x^2)$
ought to be regarded as the local term in the Hamiltonian and
$-\sigma_z^1\sigma_z^2$ as the interaction term.  

%
%
However, the same model Hamiltonian may describe many quite different
physical systems, and it is not at all clear that the splitting into
local and interaction Hamiltonians will necessarily be the same for
all these physical systems.  \emph{A priori} it does not seem that the
mathematics of quantum mechanics distinguishes any special subclass of
interaction Hamiltonians, and this makes it impossible to define a
unique splitting of $H$ into local and interaction parties on purely
mathematical grounds.  More importantly, from our point of view, the
entanglement-frustration bound holds for any splitting whatsoever,
regardless of its physical (or unphysical) origin, and it is
interesting to address the question of which splitting gives the best
value for the entanglement-frustration bound.

%
%
Let us return now to the question of saturation, and to a closer
investigation of the example of the transverse Ising model considered
in the previous section.  In this example the decomposition of
$|E_0\rangle$ into eigenstates of the local Hamiltonian $H_L$ is
equivalent to the Schmidt decomposition, and the largest Schmidt
coefficient is given by $|\langle E^L_0 | E_0 \rangle |$. Furthermore
the inequality
\begin{equation}
    \label{eq:eperp-HL-eperp}
    \langle E_\perp | H_L | E_\perp \rangle \geq E^L_0 + \Delta E_\mathrm{ent}
\end{equation}
is strict because $|E_\perp\rangle = |--\rangle$ is an excitation of
\emph{both} subsystems, whereas $E^L_0 + \Delta E_\mathrm{ent}$ is
the energy of a \emph{single} excited subsystem. The excess is
therefore the energy gap of the remaining subsystem.

\begin{figure}
\resizebox{3.0in}{!}{\includegraphics{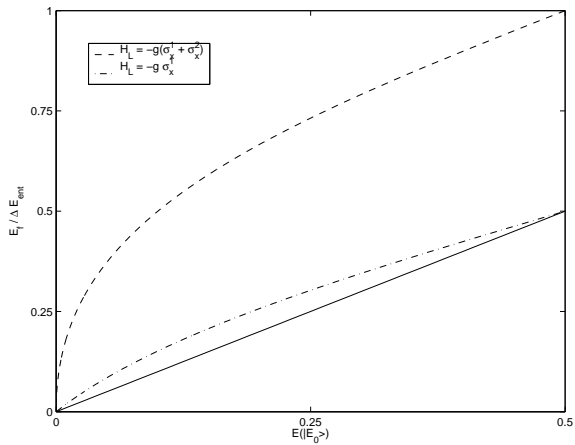}}
\caption{\label{fig:bound-comparison} Comparison of the ground-state 
  entanglement and entanglement-frustration bounds for two choices of
  splitting in the transverse Ising model. The solid line denotes the
  ideal case of saturation.}
\end{figure}

On the other hand if we take advantage of the possibility of different
splittings of $H$ to choose $H_L = -g \sigma_x^1$ then there is zero
energy associated with an excitation of the second subsystem and
Eq.~(\ref{eq:eperp-HL-eperp}) becomes an equality. The interaction
Hamiltonian is determined by the choice of local Hamiltonian, $H_I =
H-H_L = -g \sigma_x^2 - \sigma_z^1 \sigma_z^2$, and we calculate a
second entanglement-frustration bound
\begin{equation}
    \label{eq:ising-fb2}
    \frac{E_f}{\Delta E_\mathrm{ent}} = \frac{1}{2} - \left(\frac{\sqrt{1+4g^2} - \sqrt{1+g^2}}{2g} \right) .
\end{equation}
The two bounds Eq.~(\ref{eq:ising-fb}) and Eq.~(\ref{eq:ising-fb2})
are plotted against the ground-state entanglement in
Fig.~\ref{fig:bound-comparison}. It is clear that this second choice for
$H_L$ provides a substantially tighter bound, as we expect.


Let us generalize this example further.  Suppose $H$ is an arbitrary
bipartite Hamiltonian acting on two $d$-dimensional systems, with
ground state Schmidt decomposition
\begin{equation}
    \label{eq:gs-sd}
    |E_0\rangle = \lambda |a_0 b_0\rangle + \sum_{j=1}^{d-1} 
    \lambda_j |a_j b_j\rangle,
\end{equation}
where we have chosen labels so that $\lambda$ is the largest Schmidt
coefficient.  In order to ensure that Eq.~(\ref{eq:eperp-HL-eperp}) is
saturated we choose a splitting of $H$ with $H_L$ as follows:
\begin{equation}
    H_L = -\gamma |a_0\rangle \langle a_0 | \otimes I
\end{equation}
where $\gamma > 0$ is a parameter that will be chosen later in order
to best saturate the bound.  It is clear that Eq.~(\ref{eq:gs-sd}) is
an expansion of $|E_0\rangle$ in an energy eigenbasis of $H_L$, of the
same form as used in Eq.~\ref{eq:step-2-inter}, and thus that
\begin{eqnarray}
    \langle E_\perp | H_L | E_\perp \rangle & = & 0 = 
    E^L_0 + \Delta E_\mathrm{ent}.
\end{eqnarray}
It follows that for this choice of local Hamiltonian,
\begin{eqnarray}
  E(|E_0\rangle) = \frac{E_f}{\Delta E_\mathrm{ent}} +
  \left( \langle E_0 | H_I | E_0 \rangle - E^I_0 \right) / \Delta
E_\mathrm{ent},
\end{eqnarray}
i.e., the amount by which the entanglement exceeds the
entanglement-frustration bound is composed entirely of the second term
identified earlier in Eq~.(\ref{eq:bound-excess}).

To minimize this excess we choose $\gamma$ small and positive.
Observing that $H_I = H-H_L$ we may do perturbation theory in $\gamma$
to show:
\begin{eqnarray}
  E_0^I & = & E_0- \langle E_0|H_L|E_0\rangle +O(\gamma^2) \\
  & = & \langle E_0|H_I |E_0\rangle + O(\gamma^2).
\end{eqnarray}
where we used $H_I = H-H_L$ in the second line.  Using this fact
and the observation $\Delta E_{\mathrm{ent}} = \gamma$, we have
\begin{eqnarray}
    \frac{\langle E_0 | H_I | E_0 \rangle - E^I_0}{\Delta E_\mathrm{ent}} 
    & = & 
            O(\gamma).
\end{eqnarray}
Taking the limit as $\gamma \rightarrow 0$ we see that the
entanglement-frustration bound approaches the ground-state
entanglement.

In summary, we have shown:
\begin{proposition} {} \label{prop:saturation}
  Let $H$ be an arbitrary bipartite Hamiltonian. Then there exists a
  local Hamiltonian $H_L$ and corresponding interaction Hamiltonian
  $H_I$ such that the entanglement-frustration bound derived from the
  splitting $H = H_L + H_I$ is arbitrarily close to the ground-state
  entanglement of $H$.
\end{proposition}
This shown that, in principle, the entanglement-frustration bound may
be arbitrarily close to saturation for all possible values of the
ground-state entanglement $E(|E_0\rangle)$.  We therefore conclude
that the entanglement-frustration bound cannot be strengthened without
using more detailed knowledge of the system properties.

%
%
Our results show that saturation of the entanglement-frustration bound
is always possible with an appropriate choice of splitting.  They do
not, of course, tell us what splitting ought to be used, except in the
unusual situation where one knows virtually everything about the
ground-state already, in which case one may as well calculate the
ground-state entanglement directly.  Thus the content of
Proposition~\ref{prop:saturation} is not that we ought to expect to
calculate ground-state entanglement exactly, merely by choosing the
appropriate splitting for the Hamiltonian.  Rather,
Proposition~\ref{prop:saturation}, and the methods that lead to it,
tell us that the entanglement-frustration bound is the best possible,
and provide some physical guidance as to how to choose the splitting
into local and interaction Hamiltonians in order to achieve the best
possible values for the entanglement frustration bound.


\section{Ground-state entanglement and the ratio of interaction
  strength to the local energy scale}
\label{sec:ent-ratio}

%
%
The inequality Eq.~(\ref{eq:basic-inequality}) has a nice corollary
that is easily proved, relating the ground-state entanglement to a
ratio of the interaction strength with the local energy scale of the
system.  Suppose we define $E^I_{\rm max}$ to be the largest
eigenvalue of $H_I$, and let $E^I_{\rm tot} \equiv E^I_{\rm
  max}-E^I_0$ be the total energy scale for the interaction
Hamiltonian, i.e., the difference between the largest and the smallest
energies.  It follows that
\begin{eqnarray}
  E_0 & \leq & \langle E^L_0| H |E^L_0\rangle \\
      & = & \langle E^L_0|H_L|E^L_0\rangle + \langle E^L_0|H_I|E^L_0\rangle
      \\
      & \leq & E^L_0 + E^I_{\rm max}.
\end{eqnarray}
Rearranging this inequality we obtain $E_f \leq E^I_{\rm tot}$.
Combining with Eq.~(\ref{eq:basic-inequality}) then gives
\begin{eqnarray} \label{eq:ratio-result}
  E(|E_0\rangle) \leq \frac{E^I_{\rm tot}}{\Delta E_{\rm ent}}.
\end{eqnarray}

%
%
The inequality Eq.~(\ref{eq:ratio-result}) is an interesting result.
Intuition, experience, and perturbation theory tell us that if we
start with a local Hamiltonian and slowly turn on an interaction, the
ground-state entanglement will depend on how strong the interaction
is, compared with the local terms in the Hamiltonian, which tend to
keep the ground state unentangled.  Eq.~(\ref{eq:ratio-result}) is a
precise, completely general statement of this intuition, a statement
that holds even non-perturbatively.


\section{Higher-energy eigenstates and the ratio of interaction strength
to the local energy scale}
\label{sec:higher-energies}

%
%
In Sec.~\ref{sec:ent-ratio} we proved a bound,
Eq.~(\ref{eq:ratio-result}), quantifying the intuition that when an
interaction term is switched on in a many-body system, the
ground-state entanglement will depend on how strong the interaction is
compared with the strength of the local Hamiltonian.  Of course, a
similar intuition applies also for higher-energy eigenstates.
Unfortunately, the strategy used to prove Eq.~(\ref{eq:ratio-result})
cannot be applied directly to energy eigenstates other than the ground
and most excited states\footnote{We only proved
  Eq.~(\ref{eq:ratio-result}) for the ground state.  An analogous
  result for the most excited state may be proved by applying
  Eq.~(\ref{eq:ratio-result}) to the Hamiltonian $-H$.}.  The reason
is that the proof of Eq.~(\ref{eq:ratio-result}) relied on the
entanglement-frustration bound, Eq.~(\ref{eq:basic-inequality}), and
there is no natural analogue of this bound --- or even a definition of
frustration energy --- for states other than the ground and most
excited states.

%
%
In this section we prove a bound validating this intuition for all
energy eigenstates.  The bound is proved in two steps.  

%
%
First, suppose $A = B+C$, where $A$ and $B$ are normal matrices.  We
will prove a general \emph{eigenspace perturbation theorem} making
precise the intuition that $A$ and $B$ have similar eigenspaces when
$C$ is sufficiently small.  Our eigenspace perturbation theorem is a
variant on a celebrated theorem of linear algebra, the Davis-Kahan
theorem~\cite{Davis70a}\footnote{For an account of the Davis-Kahan
  theorem, see Theorem~VII.3.1 on page~211 of~\cite{Bhatia97a}, and
  the surrounding discussion in Chapter~VII of that work.}.

%
%
A detailed discussion of how our eigenspace perturbation theorem
compares to the Davis-Kahan theorem is given below.  Summarizing, the
major differences are that (a) our proof is simpler, (b) our
conclusions are more powerful, but (c) our hypotheses are more
specialized.  For these reasons, we believe our eigenspace
perturbation theorem is of substantial independent interest in its own
right.

%
%
The second step in the proof of the bound is to apply our eigenspace
perturbation theorem to understand how the entanglement in an energy
eigenstate depends on the relationship between the strength of the
local and the interaction Hamiltonians.

Let us begin with the eigenspace perturbation theorem.

\begin{theorem}[Eigenspace perturbation theorem] {} \label{thm:e-pert}
  Let $A, B$ and $C$ be matrices such that $A = B+C$, with $A$ and $B$
  normal matrices.  Let $a$ be an eigenvalue of $A$, and suppose $P_a$
  is any projector that projects onto some subspace of the
  corresponding eigenspace.  ($P_a$ may, for example, project onto the
  entire eigenspace.)  Let $\beta = \{ b \}$ be some subset of the
  eigenvalues of $B$, and let $Q_b$ be projectors projecting onto any
  subspaces of the corresponding eigenspaces of $B$.  Define $Q \equiv
  \sum_{b \in \beta} Q_b$.  Then
  \begin{eqnarray} \label{eq:eig-space-pert}
    |P_a Q| \leq \frac{|P_a C Q|}{\Delta_a} \leq \frac{U|C|U^\dagger}{\Delta_a},
  \end{eqnarray}
  where $S \leq T$ denotes a matrix inequality, i.e., $T-S$ is a
  positive matrix, $|S| \equiv \sqrt{S S^\dagger}$, $\Delta_a
  \equiv \min_{b \in \beta} |a-b|$ is the distance from $a$ to the set
  $\beta$, and $U$ is some unitary matrix.
\end{theorem}

%
%
The interpretation of these inequalities in terms of eigenspace
perturbation is perhaps not immediately clear.  Rather than describe
this interpretation immediately, we defer the description until after
the proof of the theorem and a discussion of how this result relates
to the Davis-Kahan theorem.

\textbf{Proof:} We begin by proving the first inequality. Multiplying
$A = B+C$ on the left by $P_a$ and on the right by $Q_b$, we obtain
$aP_aQ_b = bP_aQ_b+P_aCQ_ b$, which may be rearranged to give
\begin{eqnarray} \label{eq:pert-inter}
  P_aQ_b = \frac{P_a CQ_b}{a-b}.
\end{eqnarray}
Observe that $|P_aQ|^2 = P_a QP_a = \sum_b P_a Q_b Q_bP_a$.
Substituting Eq.~(\ref{eq:pert-inter}) and its adjoint gives
\begin{eqnarray}
  |P_aQ|^2 & = & \sum_b \frac{P_a C Q_b C^\dagger P_a}{|a-b|^2} \\
 & \leq & \sum_b \frac{P_a C Q_b C^\dagger P_a}{\Delta_a^2},
\end{eqnarray}
where we used $|a-b|^2 \geq \Delta_a^2$.  Summing out $b$ gives
\begin{eqnarray}
  |P_aQ|^2 & \leq & \frac{|P_a C Q|^2}{\Delta_a^2}.
\end{eqnarray}
The conclusion follows by using the operator monotonicity\footnote{A
  review of operator monotonicity may be found in Chapter~V
  of~\cite{Bhatia97a}.} of the square root function, i.e., the fact
that if $S \leq T$ then $\sqrt{S} \leq \sqrt{T}$.

To prove the second inequality in the statement of the theorem, it
obviously suffices to prove $|P_a CQ| \leq U|C|U^\dagger$.  Note first
that $P_a CQ C^\dagger P_a \leq P_a CC^\dagger P_a$.  But $P_a
CC^\dagger P_a$ and $C^\dagger P_a C$ are positive operators with the
same eigenvalues, so there exists a unitary $V$ such that $P_a
CC^\dagger P_a = V C^\dagger P_a C V^\dagger \leq VC^\dagger C
V^\dagger$.  Putting these observations together gives $P_a
CQC^\dagger P_a \leq VC^\dagger CV^\dagger$, from which it follows
that $P_A CQC^\dagger P_a \leq U C C^\dagger U^\dagger$ for some
unitary $U$.  The result now follows by using the operator
monotonicity of the square root function.

\textbf{QED}

%
%
The conclusion of Theorem~\ref{thm:e-pert} has a nice implication in
terms of matrix norms.  Suppose $\uin \cdot \uin$ is a \emph{unitarily
  invariant} matrix norm, i.e., $\uin U S V \uin = \uin S \uin$ for
any unitaries $U$ and $V$.  (Most of the familiar norms in common use
in quantum information, including all the $l_p$ norms, are easily
shown to be unitarily invariant.)  Using the polar decomposition we
see that $S = |S|U$ for some unitary $U$, and thus
Eq.~(\ref{eq:eig-space-pert}) implies that
\begin{eqnarray}
  \uin P_a Q\uin \leq \frac{\uin P_a C Q\uin}{\Delta_a} 
  \leq \frac{\uin C\uin}{\Delta_a},
\end{eqnarray}
for any unitarily invariant norm $\uin \cdot \uin$.

%
%
Let us compare the eigenspace perturbation theorem,
Theorem~\ref{thm:e-pert}, with the Davis-Kahan theorem.  The
Davis-Kahan theorem is as follows:
\begin{theorem}[Davis-Kahan theorem] {} \label{thm:D-K}
  Let $A, B$ and $C$ be matrices such that $A = B+C$, with $A$ and $B$
  normal matrices.  Let $\alpha$ and $\beta$ be subsets of the
  eigenvalues of $A$ and $B$, respectively.  Let $P$ (resp. $Q$)
  project onto the space spanned by all the eigenspaces of $A$ (resp.
  $B$) corresponding to elements of $\alpha$ (resp. $\beta$).  Suppose
  furthermore that $\alpha$ and $\beta$ are separated by an annulus of
  width $\delta$ in the complex plane, e.g., with $\alpha$ inside the
  annulus, and $\beta$ outside the annulus.  Then for any unitarily
  invariant norm $\uin \cdot \uin$,
  \begin{eqnarray}
    \uin PQ \uin \leq \frac{\uin P C Q \uin}{\delta} 
    \leq \frac{\uin C \uin}{\delta}.
  \end{eqnarray}
\end{theorem}

%
%
There are three interesting differences between the Davis-Kahan
theorem and Theorem~\ref{thm:e-pert}.  First, Theorem~\ref{thm:e-pert}
is more specialized than Davis-Kahan, in that it applies only for a
single eigenvalue of $A$, not for multiple eigenvalues.  We have tried
and failed to extend our proof to the more general case.  A second
difference is that Theorem~\ref{thm:e-pert} gives an operator
inequality that implies the corresponding inequalities for unitarily
invariant norms, but which is not implied by those inequalities.
Finally, our proof of Theorem~\ref{thm:e-pert} seems to be
substantially simpler than known proofs of the Davis-Kahan theorem.

%
%
To better understand how Theorems~\ref{thm:e-pert} and~\ref{thm:D-K}
relate to eigenspace perturbations, suppose that $P_a$ projects onto a
subspace $\mathcal{P}_a$ spanned by a single eigenstate $|a\rangle$ of
$A$, and $Q$ projects onto a subspace $\mathcal{Q}$ spanned by
eigenstates $|b\rangle$, $b \in \beta$. The norm $\uin P_a Q \uin$
turns out to measure the orthogonality of these two subspaces. For
example, in the special case when $Q$ is a rank-$1$ projector, $Q =
|b\rangle\langle b |$, we have
\begin{equation}
    \uin P_a Q \uin = \uin \, |a\rangle\langle a |b\rangle \langle b | \, \uin
  = |\langle a | b \rangle | \, \uin\, |a\rangle\langle b|\,\uin
\end{equation}
which is proportional to the cosine of the angle between $|a\rangle$
and $|b\rangle$. (Note that $\uin\, |a\rangle \langle b|\,\uin$ is a
constant independent of $|a\rangle$ and $|b\rangle$, due to unitary
invariance of the norm.)  Thus, Theorems~\ref{thm:e-pert}
and~\ref{thm:D-K} tell us that this cosine is very small (and thus
$|a\rangle$ and $|b\rangle$ are close to orthogonal) whenever the
ratio of the size of the perturbation $\uin C \uin$ to the distance
$\Delta_a$ is small.  It follows that provided $\uin C \uin$ is
sufficiently small, all the eigenvectors of $A$ and $B$ are nearly
orthogonal, except for a single nearly parallel eigenvector.

%
%
More generally, the singular values of $P_a Q$ are the cosines of what
are known as the \emph{canonical angles} between the subspaces
$\mathcal{P}$ and $\mathcal{Q}$\footnote{For an introduction to the
  canonical angles, see Chapter~VII of~\cite{Bhatia97a}, especially
  the first section.  We do not need to use any properties of the
  canonical angles in this paper.}.  If $\uin P_a Q \uin$ is small
then the cosines of the canonical angles are small, and it can be
shown that all vectors in $\mathcal{P}$ are very nearly orthogonal to
all vectors in $\mathcal{Q}$.

%
%
Let us return now to the problem of bounding the entanglement in an
arbitrary eigenstate $|E_j\rangle$ of a many-body Hamiltonian $H$. $H$
is split into a local part, $H_L$, and an interaction part, $H_I$, as
before. Our starting point is again the expansion of $|E_j\rangle$ in
terms of the eigenstates $|E^L_k\rangle$ of $H_L$. Associated to any
local Hamiltonian we can identify some natural subspaces that contain
no entanglement. These subspaces are spanned by a set of eigenstates
$|E^L_m\rangle$ related to each other by \emph{excitations or
  de-excitations of a single subsystem}. Any superposition of such
states factors into a product state, and for convenience we will refer
to such a subspace as a \emph{product subspace}.  Our use of this term
should not be confused with the more general (and more common) use of
the term product subspace, to mean any vector subspace containing no
entanglement; our use of the term is specific to a particular $H_L$,
and refers to those subspaces spanned by sets of eigenstates
$|E^L_m\rangle$ which are all related by excitations or de-excitations
of a single subsystem.
 
We will see later that for each $|E_j\rangle$ there is a natural way
to choose a corresponding product subspace from the eigenstates of
$H_L$.  For now let $\mathcal{K}$ be any such product subspace and
expand $|E_j\rangle$ in the energy eigenbasis of $H_L$ as follows
\begin{equation}
\label{eq:ej-productspace}
    |E_j\rangle = \sum_{k,\, |E^L_k\rangle \in \mathcal{K}} \alpha_k |E^L_k \rangle + \gamma |E_\perp \rangle
\end{equation}
where $|E_\perp \rangle$ is orthogonal to all states in $\mathcal{K}$.
It follows from Eq.~(\ref{eq:entanglement-measure}) that
\begin{equation}
    \label{eq:ent-alphak}
    E(|E_j\rangle) \leq 1 - \sum_k |\alpha_k|^2 = |\gamma|^2.
\end{equation}
Our strategy is to apply Theorem~\ref{thm:e-pert} to obtain a bound on
$|\gamma|^2 = 1 - \sum_k |\alpha_k|^2$.

Define $P_j$ to be the projector onto $|E_j\rangle$. We're trying to
bound the amplitude squared $|\gamma|^2$ of the component of
$|E_j\rangle$ orthogonal to $\mathcal{K}$, so let $\mathcal{K_\perp}$
denote the subspace spanned by all eigenstates $|E^L_l\rangle$ of
$H_L$ not in $\mathcal{K}$ and define $Q_{\mathcal{K}_\perp}$ to be
the corresponding projector.  Theorem~\ref{thm:e-pert} implies that
\begin{equation}
    \label{eq:dk-step1}
    \uin P_j Q_{\mathcal{K}_\perp} \uin \leq \frac{\uin H_I
      \uin}{\Delta_{j,\mathcal{K}_\perp}} 
\end{equation}
where $\Delta_{j,\mathcal{K}_\perp} = \min_{|E^L_l\rangle \in
  \mathcal{K}_\perp} |E_j - E^L_l|$. Next we must show how $\uin P_j
Q_{\mathcal{K}_\perp}\uin$ is related to the entanglement
$E(|E_j\rangle)$.

It is easily seen from Eq.~(\ref{eq:ej-productspace}) that
$Q_{\mathcal{K}_\perp} |E_j\rangle = |E_\perp\rangle$ and so
\begin{eqnarray}
\uin P_j Q_{\mathcal{K}_\perp} \uin & = & \uin \, |E_j\rangle \langle E_j | Q_{\mathcal{K}_\perp} \uin \\
        & = & |\gamma| \, \uin \, |E_j\rangle \langle E_\perp | \,\uin .
\end{eqnarray}
As remarked earlier, the value of $\uin \, |v\rangle \langle w | \,
\uin$ for any normalized vectors $|v\rangle$ and $|w\rangle$ is a
constant that depends only upon the norm $\uin \cdot \uin$.  Without
loss of generality we may assume that $\uin \, |v\rangle \langle w| \,
\uin = 1$, since multiplying a unitarily invariant norm by a constant
gives another unitarily invariant norm.  We will say any norm
satisfying this condition is \emph{normalized}.  (Examples of
normalized unitarily invariant norms include the operator norm $\| A
\| = \sup_{\||v\rangle\| = 1} \| A |v\rangle \|$ and the
Hilbert-Schmidt norm $\| A \|_2 = \sqrt{\mathrm{tr} (A A^\dagger)}$.)

Assuming that $\uin \cdot \uin$ is normalized we see that 
\begin{equation}
    \uin P_j Q_{\mathcal{K}_\perp} \uin  =  |\gamma|,  
\end{equation}
and it follows from Eqs.~(\ref{eq:ent-alphak}) and~(\ref{eq:dk-step1})
that
\begin{equation}
    E(|E_j\rangle) \leq \frac{\uin H_I \uin^2}{\left(\Delta_{j,\mathcal{K}_\perp}\right)^2} .
\end{equation}
In Appendix~\ref{app:uin} we show that for any normalized unitarily
invariant norm $\uin \cdot \uin$ we have $\|S\| \leq \uin S \uin$,
where $\|\cdot\|$ is the operator norm and $S$ any operator. The 
strongest form of the bound is therefore:
\begin{equation}
    \label{eq:dk-bound}
    E(|E_j\rangle) \leq \frac{\| H_I \|^2}
    {\left(\Delta_{j,\mathcal{K}_\perp}\right)^2} .
\end{equation}

Different choices of the product subspace $\mathcal{K}$ provide us
with a different bound in Eq.~(\ref{eq:dk-bound}). Ideally we would
like to choose $\mathcal{K}$ so that the quantity
$\Delta_{j,\mathcal{K}_\perp}$ is as large as possible. If $E_j$, or a
good approximation to $E_j$, is known then we would ensure that
$\mathcal{K}$ contained $|E^L_k\rangle$ where $|E^L_k - E_j|$ is
minimal. More typically $E_j$ is unknown, and this is not possible.
However, there is still a natural way for us to choose $\mathcal{K}$.
Importantly this choice also allows us to obtain a lower bound for
$\Delta E_{j,\mathcal{K}_\perp}$ in terms of relatively simple
quantities that depend only $H_L$ and $H_I$, not on typically
difficult-to-calculate quantities associated with the total
Hamiltonian, $H$.

Let $|E^L_j\rangle$ be the $j$-th excited eigenstate of the local
Hamiltonian. We choose the product subspace $\mathcal{K}$ so that the
expression
\begin{equation}
    \Delta E_{j,\mathrm{ent}} = \min_{|E^L_k\rangle \in \mathcal{K}_\perp} \left| E^L_j - E^L_k \right|
\end{equation}
is maximized. $\Delta E_{j,\mathrm{ent}}$ is a generalization of
$\Delta E_\mathrm{ent}$ in Sec.~\ref{sec:main}, in that it is the
energy required to excite or de-excite at least two subsystems from
the state $|E^L_j\rangle$. Note that the calculation of $\Delta
E_{j,\mathrm{ent}}$ is tedious, but in principle straightforward
provided that the energy spectrum of $H_L$ is known: simply enumerate
the possible product subspaces given the spectrum of $H_L$ (a long,
but finite list), and then calculate the minimum by inspection.

Now for each $|E^L_k\rangle \in \mathcal{K}_\perp$ we have by the
triangle inequality
\begin{eqnarray}
    |E^L_k - E_j| & \geq & |E^L_k - E^L_j| - |E^L_j - E_j| \\
        & \geq & \Delta E_{j,\mathrm{ent}} - |E^L_j - E_j| .
\end{eqnarray}
Furthermore it is straightforward to show that $|E_j - E^L_j| \leq
|E^I_\mathrm{max}|$ and so
\begin{equation}
    \Delta_{j,\mathcal{K}_\perp} =
    \min_{|E^L_k\rangle \in \mathcal{K}_\perp} \left| E^L_k - E_j \right| 
    \geq \Delta E_{j,\mathrm{ent}} - |E^I_\mathrm{max}| .
\end{equation} 
Substituting into Eq.~(\ref{eq:dk-bound}) we obtain a result in terms
of the spectrum of $H_L$ and the strength of $H_I$ alone.

\begin{proposition}
  Let $H = H_L + H_I$ with $H_L$ a local Hamiltonian, and suppose
  $\Delta_{j,\mathrm{ent}} > |E^I_\mathrm{max}|$. Then the
  entanglement in the $j$th excited eigenstate $|E_j\rangle$ of $H$,
  as measured using the definition of
  Eq.~(\ref{eq:entanglement-measure}), is bounded above by:
\begin{equation}
    \label{eq:excited-entanglement-bound}
    E(|E_j\rangle) \leq \frac{\|H_I\|^2}{\left( \Delta E_{j,\mathrm{ent}} - |E^I_\mathrm{max}| \right)^2}
\end{equation}
\end{proposition}

Noting that $|E^I_\mathrm{max}| \leq \|H_I\|$ this can be restated in a
slightly weaker but perhaps more elegant form, supposing
$\Delta_{j,\mathrm{ent}} \geq \| H_I\|$:
\begin{equation} \label{eq:excited-entanglement-bound-2}
    E(|E_j\rangle) \leq \frac{\|H_I\|^2}{\left( \Delta E_{j,\mathrm{ent}} - 
        \|H_I\| \right)^2}
    = \frac{1}{\left(\frac{\Delta E_{j,\mathrm{ent}}}{\|H_I\|} - 1 \right)^2} .
\end{equation}

Eqs.~(\ref{eq:excited-entanglement-bound})
and~(\ref{eq:excited-entanglement-bound-2}) confirm and quantify our
intuition that when the non-entangled energy scale associated with
$|E^L_j\rangle$ is large compared to the strength of the interaction
Hamiltonian we expect little entanglement in the excited state
$|E_j\rangle$ of the total Hamiltonian.

%
%
Eq.~(\ref{eq:excited-entanglement-bound}) should be compared with the
earlier result Eq.~(\ref{eq:ratio-result}) for the ground-state
entanglement.  We see that the present result is equivalent to the
earlier result, except for the presence of the term
$-|E^I_\mathrm{max}|$ in the denominator of
Eq.~(\ref{eq:excited-entanglement-bound}), which makes the present
result weaker.


\section{Conclusion}
\label{sec:conclusion}

%
%

We have introduced the frustration energy $E_f$ as a measure of the
degree of frustration between local and interaction terms in the
Hamiltonian $H = H_L+H_I$ of a many-body quantum system.  This
measure, when related to a local energy scale, allowed us to derive
the entanglement-frustration bound on the ground-state entanglement in
the system. A novel feature of this bound is that it depends only on
spectral properties of the Hamiltonians $H, H_L$ and $H_I$.
Ground-state entanglement properties can therefore be easily inferred
directly from the spectra alone.

The entanglement-frustration bound has, in turn, been used to prove a
bound, Eq.~(\ref{eq:ratio-result}), relating the ground-state
entanglement to a ratio of the strength of the interactions and an
appropriate local energy scale.  This bound involves only the
eigenvalues of the local and interaction Hamiltonians, which are
typically much easier to calculate than the eigenvalues of the full
Hamiltonian, and thus this bound is more likely to be useful in
practice.  A similar bound for an arbitrary energy eigenstate is
proved in Eqs.~(\ref{eq:excited-entanglement-bound})
and~(\ref{eq:excited-entanglement-bound-2}).



Ultimately it would be useful to have many powerful general techniques
enabling us to infer ground-state entanglement properties of a
Hamiltonian by considering the interplay between its constituent
terms. This is not always easy. For example, consider the following
system of three spin-$\frac{1}{2}$ particles
\begin{equation}
    \label{eq:three-particle}
    H = g_a H_A + g_b H_B + g_c H_C + H_{AB} + H_{BC}
\end{equation}
where $A,B,C$ label the three particles. $H_A, H_B, H_C$ are local
Hamiltonians, $H_{AB}, H_{BC}$ are interaction Hamiltonians on the
appropriate subsystem, and $g_A, g_B, g_C$ control the respective
strengths of the local Hamiltonians. The bound
Eq.~(\ref{eq:ratio-result}) derived from the entanglement-frustration
bound tells us that if $g_b$ is relatively large then there is little
entanglement between particle $B$ and the rest of the system $AC$.
From this we may deduce that if there is any entanglement in the
ground state then it must be between particles $A$ and $C$. To some
extent, then, the entanglement-frustration bound allows us to
determine the distribution of entanglement. In cases where all three
local energy scales are small compared to the interactions, however,
we are unable to directly deduce anything using the techniques in this
paper.

Throughout this paper we have defined frustration to occur when it is
not possible to find a simultaneous ground state for some local and
interaction part of a Hamiltonian. This is based on an analogy to the
useful definition of frustration, which involves competition between
interactions, as discussed in the introduction, and illustrated
Fig.~\ref{fig:frustration}.  (An insightful review of classical and
quantum frustration in this sense may be found in~\cite{Wolf03a}.)
Both these points of view suggest interesting extensions of the
investigations in the present paper.

%
%
For example, we believe that quantum frustration suggests interesting
parallels with the phenomenon of \emph{entanglement
  sharing}~\cite{Coffman00a} which places restrictions on the
distribution of entanglement amongst many particles. In particular, we
expect non-trivial distributions of entanglement in the ground state
of two overlapping interactions. For example, consider a Hamiltonian
acting on three spin-$\frac{1}{2}$ particles as before
\begin{equation}
    \label{eq:three-particles2}
    H = H_{AB} + H_{BC} 
\end{equation}
and suppose that $H_{AB}$ and $H_{BC}$ have non-degenerate, maximally
entangled ground states. It is impossible for entanglement to be
distributed in a way that would provide a ground state for $H$ that is
a simultaneous ground state of $H_{AB}$ and $H_{BC}$. The system is
therefore necessarily frustrated. We might ask what happens to the
ground-state entanglement distribution in systems such as this, and
whether there are any properties of the constituent Hamiltonians that
allow us to prove quantitative bounds relating the distribution of
two-party, GHZ-type and W-type entanglement in this system.

\acknowledgments

We thank Henry Haselgrove and Guifre Vidal for enjoyable and
encouraging discussions.  Much of this work was completed at the
Institute for Quantum Information at Caltech, and we thank the
Institute for their hospitality.

\appendix
\section{Unitarily invariant norms and the sup norm}
\label{app:uin}

%
%
In this Appendix we prove that for any normalized unitarily invariant
norm $\uin \cdot \uin$, $\|S\| \leq \uin S\uin$, where $\| \cdot \|$
is the usual matrix norm.  This result is an easy corollary of the
following theorem, proved as Theorem~3.5.5 on page~204
of~\cite{Horn91a}:
\begin{theorem}
  Let $\uin \cdot \uin$ be a unitarily invariant norm on the space of
  $n \times n$ matrices.  Then there exists a compact set $M$ in
  $\Re^n$ whose elements are decreasing sequences of positive real
  numbers, and such that
  \begin{eqnarray}
    \uin S \uin = \max_{m \in M} \sum_{j=1}^n m_j \sigma_j(S),
  \end{eqnarray}
  where $\sigma_j(S)$ are the singular values of $S$, arranged into
  decreasing order.
\end{theorem}

%
%
For a normalized unitarily invariant norm we have $\uin \, |s\rangle
\langle s| \, \uin = 1$, where $|s\rangle$ is any normalized state.
It follows from the theorem that there exists $m \in M$ such that $m_1
= 1$.  It follows that for any $S$, $\uin S \uin \geq \sigma_1(S) = \|
S\|$, as we set out to prove.

\bibliography{../mybib}

\begin{thebibliography}{32}
\expandafter\ifx\csname natexlab\endcsname\relax\def\natexlab#1{#1}\fi
\expandafter\ifx\csname bibnamefont\endcsname\relax
  \def\bibnamefont#1{#1}\fi
\expandafter\ifx\csname bibfnamefont\endcsname\relax
  \def\bibfnamefont#1{#1}\fi
\expandafter\ifx\csname citenamefont\endcsname\relax
  \def\citenamefont#1{#1}\fi
\expandafter\ifx\csname url\endcsname\relax
  \def\url#1{\texttt{#1}}\fi
\expandafter\ifx\csname urlprefix\endcsname\relax\def\urlprefix{URL }\fi
\providecommand{\bibinfo}[2]{#2}
\providecommand{\eprint}[2][]{\url{#2}}

\bibitem[{\citenamefont{Haselgrove
  et~al.}(2003{\natexlab{a}})\citenamefont{Haselgrove, Nielsen, and
  Osborne}}]{Haselgrove03c}
\bibinfo{author}{\bibfnamefont{H.~L.} \bibnamefont{Haselgrove}},
  \bibinfo{author}{\bibfnamefont{M.~A.} \bibnamefont{Nielsen}},
  \bibnamefont{and} \bibinfo{author}{\bibfnamefont{T.~J.}
  \bibnamefont{Osborne}}, \bibinfo{journal}{{arXiv}:quant-ph/0308083}
  (\bibinfo{year}{2003}{\natexlab{a}}).

\bibitem[{\citenamefont{Haselgrove
  et~al.}(2003{\natexlab{b}})\citenamefont{Haselgrove, Nielsen, and
  Osborne}}]{Haselgrove03b}
\bibinfo{author}{\bibfnamefont{H.~L.} \bibnamefont{Haselgrove}},
  \bibinfo{author}{\bibfnamefont{M.~A.} \bibnamefont{Nielsen}},
  \bibnamefont{and} \bibinfo{author}{\bibfnamefont{T.~J.}
  \bibnamefont{Osborne}}, \bibinfo{journal}{Phys. Rev. Lett.}
  \textbf{\bibinfo{volume}{91}}, \bibinfo{pages}{052311}
  (\bibinfo{year}{2003}{\natexlab{b}}),
  \bibinfo{note}{{arXiv}:quant-ph/0303022}.

\bibitem[{\citenamefont{Vidal et~al.}(2003)\citenamefont{Vidal, Latorre, Rico,
  and Kitaev}}]{Vidal03a}
\bibinfo{author}{\bibfnamefont{G.}~\bibnamefont{Vidal}},
  \bibinfo{author}{\bibfnamefont{J.~I.} \bibnamefont{Latorre}},
  \bibinfo{author}{\bibfnamefont{E.}~\bibnamefont{Rico}}, \bibnamefont{and}
  \bibinfo{author}{\bibfnamefont{A.}~\bibnamefont{Kitaev}},
  \bibinfo{journal}{Phys. Rev. Lett.} \textbf{\bibinfo{volume}{90}},
  \bibinfo{pages}{227902} (\bibinfo{year}{2003}),
  \bibinfo{note}{{arXiv}:quant-ph/0211074}.

\bibitem[{\citenamefont{Latorre et~al.}(2003)\citenamefont{Latorre, Rico, and
  Vidal}}]{Latorre03a}
\bibinfo{author}{\bibfnamefont{J.~I.} \bibnamefont{Latorre}},
  \bibinfo{author}{\bibfnamefont{E.}~\bibnamefont{Rico}}, \bibnamefont{and}
  \bibinfo{author}{\bibfnamefont{G.}~\bibnamefont{Vidal}},
  \bibinfo{journal}{{arXiv}:quant-ph/0304098}  (\bibinfo{year}{2003}).

\bibitem[{\citenamefont{Tessier et~al.}(2003)\citenamefont{Tessier, Deutsch,
  Delgado, and Fuentes-Guridi}}]{Tessier03a}
\bibinfo{author}{\bibfnamefont{T.}~\bibnamefont{Tessier}},
  \bibinfo{author}{\bibfnamefont{I.~H.} \bibnamefont{Deutsch}},
  \bibinfo{author}{\bibfnamefont{A.}~\bibnamefont{Delgado}}, \bibnamefont{and}
  \bibinfo{author}{\bibfnamefont{I.}~\bibnamefont{Fuentes-Guridi}},
  \bibinfo{journal}{{arXiv}:quant-ph/0306015}  (\bibinfo{year}{2003}).

\bibitem[{\citenamefont{Costi and McKenzie}(2003)}]{Costi03a}
\bibinfo{author}{\bibfnamefont{T.~A.} \bibnamefont{Costi}} \bibnamefont{and}
  \bibinfo{author}{\bibfnamefont{R.~H.} \bibnamefont{McKenzie}},
  \bibinfo{journal}{Phys. Rev. A} \textbf{\bibinfo{volume}{68}},
  \bibinfo{pages}{034301} (\bibinfo{year}{2003}),
  \bibinfo{note}{{arXiv}:quant-ph/0302055}.

\bibitem[{\citenamefont{Hines et~al.}(2003)\citenamefont{Hines, McKenzie, and
  Milburn}}]{Hines03a}
\bibinfo{author}{\bibfnamefont{A.~P.} \bibnamefont{Hines}},
  \bibinfo{author}{\bibfnamefont{R.~H.} \bibnamefont{McKenzie}},
  \bibnamefont{and} \bibinfo{author}{\bibfnamefont{G.~J.}
  \bibnamefont{Milburn}}, \bibinfo{journal}{Phys. Rev. A}
  \textbf{\bibinfo{volume}{67}}, \bibinfo{pages}{013609}
  (\bibinfo{year}{2003}), \bibinfo{note}{{arXiv}:quant-ph/0209122}.

\bibitem[{\citenamefont{Osterloh et~al.}(2002)\citenamefont{Osterloh, Amico,
  Falci, and Fazio}}]{Osterloh02a}
\bibinfo{author}{\bibfnamefont{A.}~\bibnamefont{Osterloh}},
  \bibinfo{author}{\bibfnamefont{L.}~\bibnamefont{Amico}},
  \bibinfo{author}{\bibfnamefont{G.}~\bibnamefont{Falci}}, \bibnamefont{and}
  \bibinfo{author}{\bibfnamefont{R.}~\bibnamefont{Fazio}},
  \bibinfo{journal}{Nature} \textbf{\bibinfo{volume}{416}},
  \bibinfo{pages}{608} (\bibinfo{year}{2002}),
  \bibinfo{note}{{arXiv}:quant-ph/0202029}.

\bibitem[{\citenamefont{Osborne and Nielsen}(2002)}]{Osborne02a}
\bibinfo{author}{\bibfnamefont{T.~J.} \bibnamefont{Osborne}} \bibnamefont{and}
  \bibinfo{author}{\bibfnamefont{M.~A.} \bibnamefont{Nielsen}},
  \bibinfo{journal}{Phys. Rev. A} \textbf{\bibinfo{volume}{66}},
  \bibinfo{pages}{032110} (\bibinfo{year}{2002}),
  \bibinfo{note}{{arXiv}:quant-ph/0202162}.

\bibitem[{\citenamefont{Scheel et~al.}(2002)\citenamefont{Scheel, Eisert,
  Knight, and Plenio}}]{Scheel02a}
\bibinfo{author}{\bibfnamefont{S.}~\bibnamefont{Scheel}},
  \bibinfo{author}{\bibfnamefont{J.}~\bibnamefont{Eisert}},
  \bibinfo{author}{\bibfnamefont{P.~L.} \bibnamefont{Knight}},
  \bibnamefont{and} \bibinfo{author}{\bibfnamefont{M.~B.}
  \bibnamefont{Plenio}}, \bibinfo{journal}{J. Mod. Opt.}
  \textbf{\bibinfo{volume}{50}}, \bibinfo{pages}{881} (\bibinfo{year}{2002}),
  \bibinfo{note}{{arXiv}:quant-ph/0207120}.

\bibitem[{\citenamefont{Wang and Zanardi}(2002)}]{Wang02b}
\bibinfo{author}{\bibfnamefont{X.}~\bibnamefont{Wang}} \bibnamefont{and}
  \bibinfo{author}{\bibfnamefont{P.}~\bibnamefont{Zanardi}},
  \bibinfo{journal}{Phys.~Lett.~A} \textbf{\bibinfo{volume}{301}},
  \bibinfo{pages}{1} (\bibinfo{year}{2002}),
  \bibinfo{note}{{arXiv}:quant-ph/0202108}.

\bibitem[{\citenamefont{O'Connor and Wootters}(2001)}]{Oconnor01a}
\bibinfo{author}{\bibfnamefont{K.~M.} \bibnamefont{O'Connor}} \bibnamefont{and}
  \bibinfo{author}{\bibfnamefont{W.~K.} \bibnamefont{Wootters}},
  \bibinfo{journal}{Phys. Rev. A} \textbf{\bibinfo{volume}{63}},
  \bibinfo{pages}{052302} (\bibinfo{year}{2001}), \bibinfo{note}{0009041}.

\bibitem[{\citenamefont{Gunlycke et~al.}(2001)\citenamefont{Gunlycke, Kendon,
  Vedral, and Bose}}]{Gunlycke01a}
\bibinfo{author}{\bibfnamefont{D.}~\bibnamefont{Gunlycke}},
  \bibinfo{author}{\bibfnamefont{V.~M.} \bibnamefont{Kendon}},
  \bibinfo{author}{\bibfnamefont{V.}~\bibnamefont{Vedral}}, \bibnamefont{and}
  \bibinfo{author}{\bibfnamefont{S.}~\bibnamefont{Bose}},
  \bibinfo{journal}{Phys. Rev. A} \textbf{\bibinfo{volume}{64}},
  \bibinfo{pages}{042302} (\bibinfo{year}{2001}),
  \bibinfo{note}{{arXiv}:quant-ph/0102137}.

\bibitem[{\citenamefont{Nielsen}(1998)}]{Nielsen98d}
\bibinfo{author}{\bibfnamefont{M.~A.} \bibnamefont{Nielsen}}, Ph.D. thesis,
  \bibinfo{school}{University of New Mexico} (\bibinfo{year}{1998}),
  \bibinfo{note}{{arXiv}:quant-ph/0011036}.

\bibitem[{\citenamefont{Jordan and B\"{u}ttiker}(2003)}]{Jordan03a}
\bibinfo{author}{\bibfnamefont{A.~N.} \bibnamefont{Jordan}} \bibnamefont{and}
  \bibinfo{author}{\bibfnamefont{M.}~\bibnamefont{B\"{u}ttiker}}
  (\bibinfo{year}{2003}), \bibinfo{note}{{arXiv}:quant-ph/0311647}.

\bibitem[{\citenamefont{Nielsen and Chuang}(2000)}]{Nielsen00a}
\bibinfo{author}{\bibfnamefont{M.~A.} \bibnamefont{Nielsen}} \bibnamefont{and}
  \bibinfo{author}{\bibfnamefont{I.~L.} \bibnamefont{Chuang}},
  \emph{\bibinfo{title}{Quantum computation and quantum information}}
  (\bibinfo{publisher}{Cambridge University Press},
  \bibinfo{address}{Cambridge}, \bibinfo{year}{2000}).

\bibitem[{\citenamefont{Preskill}(1998)}]{Preskill98c}
\bibinfo{author}{\bibfnamefont{J.}~\bibnamefont{Preskill}},
  \emph{\bibinfo{title}{Physics 229: Advanced mathematical methods of physics
  --- Quantum computation and information}} (\bibinfo{publisher}{California
  Institute of Technology}, \bibinfo{address}{Pasadena, CA},
  \bibinfo{year}{1998}),
  \bibinfo{note}{http://www.theory.caltech.edu/people/preskill/ph229/}.

\bibitem[{\citenamefont{Aeppli and Chandra}(1997)}]{Aeppli97a}
\bibinfo{author}{\bibfnamefont{G.}~\bibnamefont{Aeppli}} \bibnamefont{and}
  \bibinfo{author}{\bibfnamefont{P.}~\bibnamefont{Chandra}},
  \bibinfo{journal}{Science} \textbf{\bibinfo{volume}{275}},
  \bibinfo{pages}{177} (\bibinfo{year}{1997}).

\bibitem[{\citenamefont{Nielsen}(2002)}]{Nielsen02e}
\bibinfo{author}{\bibfnamefont{M.~A.} \bibnamefont{Nielsen}},
  \bibinfo{journal}{Sci. Am.} \textbf{\bibinfo{volume}{287}},
  \bibinfo{pages}{66} (\bibinfo{year}{2002}).

\bibitem[{\citenamefont{Vidal}(2003)}]{Vidal03b}
\bibinfo{author}{\bibfnamefont{G.}~\bibnamefont{Vidal}},
  \bibinfo{journal}{Phys. Rev. Lett.} \textbf{\bibinfo{volume}{91}},
  \bibinfo{pages}{147902} (\bibinfo{year}{2003}),
  \bibinfo{note}{{arXiv}:quant-ph/0301063}.

\bibitem[{\citenamefont{Osborne}(2002)}]{Osborne02b}
\bibinfo{author}{\bibfnamefont{T.~J.} \bibnamefont{Osborne}}, Ph.D. thesis,
  \bibinfo{school}{The University of Queensland} (\bibinfo{year}{2002}).

\bibitem[{\citenamefont{Preskill}(2000)}]{Preskill00a}
\bibinfo{author}{\bibfnamefont{J.}~\bibnamefont{Preskill}},
  \bibinfo{journal}{J. Mod. Opt.} \textbf{\bibinfo{volume}{47}},
  \bibinfo{pages}{127} (\bibinfo{year}{2000}),
  \bibinfo{note}{{arXiv}:quant-ph/9904022}.

\bibitem[{\citenamefont{Horodecki}(2001)}]{Horodecki01a}
\bibinfo{author}{\bibfnamefont{M.}~\bibnamefont{Horodecki}},
  \bibinfo{journal}{Quantum Information and Computation}
  \textbf{\bibinfo{volume}{1}}, \bibinfo{pages}{3} (\bibinfo{year}{2001}).

\bibitem[{\citenamefont{Terhal}(2002)}]{Terhal02a}
\bibinfo{author}{\bibfnamefont{B.~M.} \bibnamefont{Terhal}},
  \bibinfo{journal}{Theoretical Computer Science}
  \textbf{\bibinfo{volume}{287}}, \bibinfo{pages}{313} (\bibinfo{year}{2002}),
  \bibinfo{note}{{arXiv}:quant-ph/0101032}.

\bibitem[{\citenamefont{Vedral et~al.}(1997)\citenamefont{Vedral, Plenio,
  Rippin, and Knight}}]{Vedral97a}
\bibinfo{author}{\bibfnamefont{V.}~\bibnamefont{Vedral}},
  \bibinfo{author}{\bibfnamefont{M.~B.} \bibnamefont{Plenio}},
  \bibinfo{author}{\bibfnamefont{M.~A.} \bibnamefont{Rippin}},
  \bibnamefont{and} \bibinfo{author}{\bibfnamefont{P.~L.}
  \bibnamefont{Knight}}, \bibinfo{journal}{Phys. Rev. Lett.}
  \textbf{\bibinfo{volume}{78}}, \bibinfo{pages}{2275} (\bibinfo{year}{1997}),
  \bibinfo{note}{{arXiv}:quant-ph/9702027}.

\bibitem[{\citenamefont{Vedral and Plenio}(1998)}]{Vedral98a}
\bibinfo{author}{\bibfnamefont{V.}~\bibnamefont{Vedral}} \bibnamefont{and}
  \bibinfo{author}{\bibfnamefont{M.~B.} \bibnamefont{Plenio}},
  \bibinfo{journal}{Phys. Rev. A} \textbf{\bibinfo{volume}{57}},
  \bibinfo{pages}{1619} (\bibinfo{year}{1998}).

\bibitem[{\citenamefont{Biham et~al.}(2002)\citenamefont{Biham, Nielsen, and
  Osborne}}]{Biham02a}
\bibinfo{author}{\bibfnamefont{O.}~\bibnamefont{Biham}},
  \bibinfo{author}{\bibfnamefont{M.~A.} \bibnamefont{Nielsen}},
  \bibnamefont{and} \bibinfo{author}{\bibfnamefont{T.~J.}
  \bibnamefont{Osborne}}, \bibinfo{journal}{Phys. Rev. A}
  \textbf{\bibinfo{volume}{65}}, \bibinfo{pages}{062312}
  (\bibinfo{year}{2002}).

\bibitem[{\citenamefont{Davis and Kahan}(1970)}]{Davis70a}
\bibinfo{author}{\bibfnamefont{C.}~\bibnamefont{Davis}} \bibnamefont{and}
  \bibinfo{author}{\bibfnamefont{W.~M.} \bibnamefont{Kahan}},
  \bibinfo{journal}{SIAM J. Numer. Anal.} \textbf{\bibinfo{volume}{7}},
  \bibinfo{pages}{1} (\bibinfo{year}{1970}).

\bibitem[{\citenamefont{Bhatia}(1997)}]{Bhatia97a}
\bibinfo{author}{\bibfnamefont{R.}~\bibnamefont{Bhatia}},
  \emph{\bibinfo{title}{Matrix analysis}}
  (\bibinfo{publisher}{Springer-Verlag}, \bibinfo{address}{New York},
  \bibinfo{year}{1997}).

\bibitem[{\citenamefont{Wolf et~al.}(2003)\citenamefont{Wolf, Verstaete, and
  Cirac}}]{Wolf03a}
\bibinfo{author}{\bibfnamefont{M.~M.} \bibnamefont{Wolf}},
  \bibinfo{author}{\bibfnamefont{F.}~\bibnamefont{Verstaete}},
  \bibnamefont{and} \bibinfo{author}{\bibfnamefont{J.~I.} \bibnamefont{Cirac}}
  (\bibinfo{year}{2003}), \bibinfo{note}{{arXiv}:quant-ph/0311051}.

\bibitem[{\citenamefont{Coffman et~al.}(2000)\citenamefont{Coffman, Kundu, and
  Wootters}}]{Coffman00a}
\bibinfo{author}{\bibfnamefont{V.}~\bibnamefont{Coffman}},
  \bibinfo{author}{\bibfnamefont{J.}~\bibnamefont{Kundu}}, \bibnamefont{and}
  \bibinfo{author}{\bibfnamefont{W.~K.} \bibnamefont{Wootters}},
  \bibinfo{journal}{Phys. Rev. A} \textbf{\bibinfo{volume}{61}},
  \bibinfo{pages}{052306} (\bibinfo{year}{2000}),
  \bibinfo{note}{{arXiv}:quant-ph/9907047}.

\bibitem[{\citenamefont{Horn and Johnson}(1991)}]{Horn91a}
\bibinfo{author}{\bibfnamefont{R.~A.} \bibnamefont{Horn}} \bibnamefont{and}
  \bibinfo{author}{\bibfnamefont{C.~R.} \bibnamefont{Johnson}},
  \emph{\bibinfo{title}{Topics in matrix analysis}}
  (\bibinfo{publisher}{Cambridge University Press},
  \bibinfo{address}{Cambridge}, \bibinfo{year}{1991}).

\end{thebibliography}

\end{document}